# Jensenone from eucalyptus essential oil as a potential inhibitor of COVID 19 corona virus infection

*Arun Dev Sharma\*, Inderjeet Kaur*

*PG dept of Biotechnology, Lyallpur Khalsa College Jalandhar*

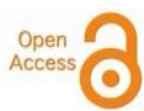

**Abstract**

COVID-19, a member of corona virus family is spreading its tentacles across the world due to lack of drugs at present. However, the main viral proteinase (Mpro/3CLpro) has recently been regarded as a suitable target for drug design against SARS infection due to its vital role in polyproteins processing necessary for coronavirus reproduction. The present in silico study was designed to evaluate the effect of Jensenone, a essential oil component from eucalyptus oil, on Mpro by docking study. In the present study, molecular docking studies were conducted by using 1-click dock and swiss dock tools. Protein interaction mode was calculated by Protein Interactions Calculator. The calculated parameters such as binding energy, and binding site similarity indicated effective binding of Jensenone to COVID-19 proteinase. Active site prediction further validated the role of active site residues in ligand binding. PIC results indicated that, Mpro/ Jensenone complexes forms hydrophobic interactions, hydrogen bond interactions and strong ionic interactions. Therefore, Jensenone may represent potential treatment potential to act as COVID-19 Mpro inhibitor. However, further research is necessary to investigate their potential medicinal use.

**Graphical abstract**

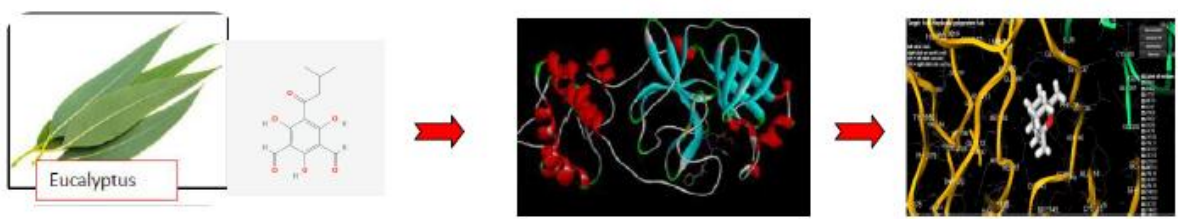

1. Virtual ligand, Jensenone    2. Mpro protein of COVID-19    3. Molecular docking

## Introduction

COVID-19 is easily transmissible and it has already been spread worldwide. Symptoms are flu-like and can include fever, muscle and body aches, coughing, and sore throat. Symptoms may appear 5-6 days after infection. As of March 20h, 2020, over 243,000 cases of COVID-19 have been confirmed worldwide, over 10,000 of which have resulted in death. At present, no specific therapies for COVID-19 are available and research regarding the treatment of COVID-19 are infancy. However, the measures that have been implemented remain limited to preventive and supportive therapies, designed to prevent further complications and organ damage [Morales et al., 2020]. Some preliminary studies have investigated potential combinations that include anti malarial drug chlorouinone, and anti-HIV vaccines can be used to treat COVID-19 infections. A separate investigation performed by Lu (2020) indicated that among 4 tested drugs (nelfinavir, pitavastatin, perampanel, and praziquantel), nelfinavir was identified as the best potential inhibitor against COVID-19.

The main protease (Mpro)/chymotrypsin-like protease (3CLpro) from COVID-19, represents a potential target for the inhibition of CoV replication [Lu, 2020]. It was observed that genome of CoV encodes two proteins ppla and pplb which are involved in spike, membrane, envelop, nucleoprotein, replicase, and polymerase activity of viruses. This function is performed by main protease (Mpro/3CLpro) (Liu and Wang, 2020). The Mpro has 3 structural domains; domain I (residues 8 - 101) and domain II (residues 102 - 184) both have beta barrel motifs representing chymotrypsin catalytic domain and domain III (residues 185 - 200) with a helical structure participates in dimerization of protein and active enzyme production. Given its vital role in polyprotein processing and virus maturation, Mpro is considered to be a suitable target for viral inhibitor development as an approach toward SARS.

Environmental factors can greatly influence the secretion of secondary metabolites from aromatic plants. Therefore, great attention has been paid to the secondary metabolites secreted by plants that may be developed as medicines [Xang et al., 2018]. Among several compounds, essential oils, from medicinal plants, have been reported to have antiviral bioactivities [Zakaryan et al., 2017]. Essential oils are highly volatile substances in a mixture of terpenes, oxygenated derivatives and other aromatic compounds. Jensenone is the another component found in eucalyptus oil from all eucalyptus plants (Goodger et al., 2016). Jensenone is a natural organic compound that is a colorless liquid. It is a cyclic ether and a monoterpenoid. In the present study, we investigated Jensenone as potential inhibitor candidates for COVID-19 Mpro. The findings of the present study will provide other researchers with opportunities to identify the right drug to combat COVID-19.

**Material and methods**

*Proteins/Macromolecules*

COVID-19 3CLpro/Mpro structures were obtained from PDB (https://www.rcsb.org/). The native ligand for 3clpro/Mpro structures was Jensenone.

*Ligand and Drug Scan*

The 3-dimensional (3D) structure of Jensenone was obtained from PubChem (https://pubchem.ncbi.nlm.nih.gov/). PubChem is a chemical substance and biological activities repository consisting of three databases, including substance, compound, and bioassay databases.

*Binding Mode of Docked Complexes*

The docked complex structure output format was submitted into the Protein Interactions Calculator (PIC) webserver

(http://pic.mbu.iisc .ernet.in/) in order to map the interaction of the resulting docked complex. The parameters such as number of hydrogen bonds, number of hydrophobic residues, and number of aromatic and ionic interactions were considered in interpreting the strength of the interaction

*Determination of Active Sites*

The amino acids in the active site of a protein were determined using the Computed Atlas for Surface Topography of Proteins (CASTp) (http://sts.bioe.uic.edu/castp/index.html?201l).

*Molecular Docking*

The docking study of the compound over COVID-19 Mpro was studied using 1-click docking (https://mcule.com/apps/1-click-docking/) and swiss doc (http://www.swissdock.ch/ ) softwares.

### Table 1: Properties of COVID-19 Mpro potential inhibitor Jensenone

| Ligand | Property | Value |
|---|---|---|
| 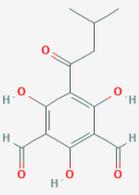 | Mass | 266.2459 |
| | logP | 1.6572 |
| | H-bond acceptors | 6 |
| | H-bond donors | 3 |
| | Rotatable bonds | 5 |
| | PSA | 111.9000 |
| | RO5 violations | 0 |
| | RO3 violations | 3 |
| | Refractivity | 67.9015 |
| | Atoms | 33 |
| | Rings | 1 |
| | Heavy atoms | 19 |
| | Hydrogen atoms | 14 |

| Ligand | Property | Value |
|---|---|---|
| | Heteroatoms | 6 |
| | N/O atoms | 6 |
| | Inorganic atoms | 0 |
| | Halogen atoms | 0 |
| | Chiral centers | 0 |
| | R/S chiral centers | 0 |
| | Unknown chiral centers | 0 |
| | Undefined chiral centers | 0 |
| | Stereo double bonds | 0 |
| | Cis/trans stereo double bonds | 0 |
| | Unknown stereo double bonds | 0 |
| | Undefined stereo double bonds | 0 |

**Result and Discussion**

Computational and structural biology methods have accelerated the discovery of novel drugs used to treat viral diseases. Corona viruses (CoVs) are a group of viruses that infect animals and humans. CoV infections affect animals in various ways like: respiratory, fever, cold, digestive, and liver systems of humans and animals. It is the main protease (Mpro) found in the CoV associated with the severe acute respiratory syndrome (SARS), which can be accessed in PDB and was suggested to be a potential drug target for 2019-nCov [Lu, 2020]. In many viruses, proteases play essential roles in viral replication; therefore, proteases are often used as protein targets during the development of antiviral therapeutics. In CoV, the Mpro protein is involved in virus proteolytic maturation and has been examined as a potential target protein by inhibiting the cleavage of the viral polyprotein to prevent the spread of infection. The invention of the Mpro/3CLpro protease structure in COVID-19 provides a nice path to identify potential drug candidates to prevent infection. As cited by Liu and Wang (2020), **p**roteases represent key targets for the inhibition virus replication, and the protein sequences of the SARS-CoV Mpro and the 2019-nCoV Mpro are 96% identical, hence host proteases can be used as potential therapeutic targets. We followed the structural biology aspects which focus on the availability and retrieval of a main protease (Mpro) or 3C-like protease (3CLpro) receptor structure from PDB database. The ligand (Jensenone) was docked to main protease (Mpro) or 3C-like protease (3CLpro) using 1-click and swiss dock softwares. Table 1 and 2 shows the structure of ligand and amino acids found in the active site pockets of Mpro.

**Table 2: Protein target structure, native ligand and active site amino acids**

| Pdb id | Macromolecule | Native ligand | Active site |
|---|---|---|---|
| 1UK3 | 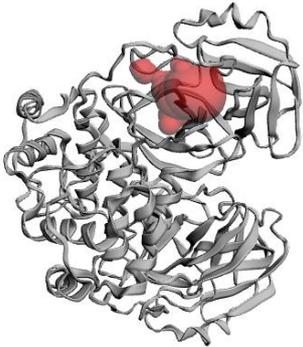 | 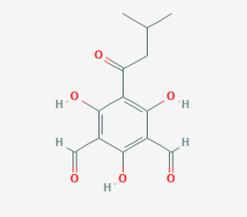 | PHE18,LEU139,ASN140, GLY141,SER142,CYS143, PRO166 |

Previously several ligands and drug candidate compounds have been selected, as per criteria of Lipinski's rule of five. So the ligands that did not incur more than 2 violations of Lipinski's rule could be used in molecular docking experiments with the target protein. The drug scanning results (Table 1) show that Jensenone, ligand used in this study, was accepted by Lipinski's rule of five. 1-click docking tool that was used to find out interaction of inhibitor i.e Jensenone with Mpro/3CLpro protein revealed 4 different poses based on the dock score and the pose with highest negative values indicated maximum binding affinity (Table 3). The results of Swiss Dock are shown in Table 3. The results of Swiss Dock showed full fitness and Gibbs free energy predicts docking interactions. The Jensenone showed full fitness of -2290.07 kcal/mol and estimated ΔG of -6.03 kcal/mol within active site amino acids PHE138 and HIS161 of

**Table 3. Binding Energy and full fitness values of Docked complex**

| Inhibitor | Dock pose | Dock score (-) |
|---|---|---|
| Jensenone | 1 | 5.5 |
|  | 2 | 5.5 |
|  | 3 | 5.0 |
|  | 4 | 4.8 |

| Inhibitor | No of swiss doc Clusters | Full fitness (kcalmol)(-) | Estimated Δ G (kcalmol) (-) | Cluster rank | |
|---|---|---|---|---|---|
| Jensenone | 38 | 2290.07 | 6.03 | 0 | 0 |
| | | 2290.06 | 6.02 | 0 | 1 |
| | | 2290.02 | 6.01 | 0 | 2 |
| | | 2289.93 | 6.00 | 0 | 3 |
| | | 2289.92 | 5.93 | 0 | 4 |
| | | 2289.74 | 5.91 | 0 | 5 |

Mpro/3CLpro proteins of COVID-19 (Fig. 1). Located at the interface between domains I and II, the 2 conserved residues His61 and PHE138 form the catalytic dyad of Mpro. CAST-P server also revealed the presence of PHE138 as active site residue of cavity 3 in Mpro protein (Table 1). Several compounds, such as flavonoids, terpenoids and phenolics from essential oils, have been reported to show antiviral bioactivities [Im et al., 2015; Zakarayan et al., 2017]. We investigated that jensenone as potential inhibitor of the COVID-19 Mpro. Hydroxy groups (-OH), ketone groups (=O) and ether groups (-O-) in Jensenone compounds are predicted to play roles amino acid residue interactions at the active site of COVID-19 Mpro.

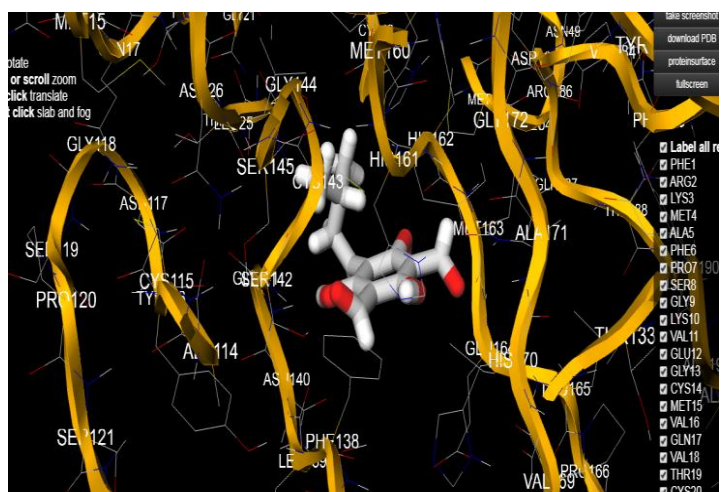

*Fig 1: Representation of docking of Jensenone with COVID-19 Mpro protein*

Once we observed that the Jensenone could potentially bind to Mpro/3CLpro protein, the next step was to know the binding mode. In particular, we have used Protein Interactions Calculator (PIC) to recognize the interactions within the bound complexes. In structural bioinformatics, predicting protein-protein interactions which stabilize the tertiary and quaternary structures is an important task. For the top best four Mpro/ Jensenone complexes with the best cluster size were subjected to PIC server and the binding mode (interactions) of each peptide are given in Table 2. PIC identified interactions such as hydrophobic residues interactions, ionic interactions, hydrogen bonds, aromatic-aromatic interactions and aromatic–sulphur interactions within the peptide-protein complexes. According to the PIC server results as shown in Table 4, Mpro/ Jensenone complexes forms hydrophobic interactions with ALA7,PRO52,TRP207,LEU29,TRY126, PRO184; hydrogen bond interactions with M4,V18,L30,D10,T16; and ionic interactions with LYS3,ASP34,ARG38,HIS163 as shown in Table 4. These residues may be considered as key or critical and may play a major role in the protein protein-interaction and might inhibit the formation of the beta-barrel motif of Mpro/3CLpro. Further studies may help to understand the role of these residues in drug binding mechanism. Structural flexibility is one of the important physical properties that affect protein conformation and function. Whereas high increase in kinetics energy and protein flexibility can disrupt non covalent interactions as in thermal denaturation; a sharp decrease in flexibility can also cause protein denaturation as seen in cold denaturation. Therefore, proteins need an essential amount of flexibility to carry out their native function at physiological conditions. In this context, an inhibitor by binding to a protein can alter its flexibility and decrease its enzymatic activity. Finally, lack of wet-lab validation is a drawback in our research and we expect computational biology analysis and its integration with wet-lab data can be productive in the determination of potential anti-Mpro/3CLpro components.

**Table 4: Binding mode of each peptide-protein complex using Protein Interaction Calculator (PIC) server.**

| Hydrophobic interactions | Main chain Hydrogen bond interactions | Main chain-Side chain Hydrogen bond interactions | Side chain-Side chain Hydrogen bond interactions | Intra protein Ionic interactions | Aromatic aromatic interactions | Aromatic sulphur interactions | Intraprotein Cation-Pi Interactions |
|---|---|---|---|---|---|---|---|
| PRO7 | M4 | PHE6 | Ser8 | LYS3 | PHE1 | TYR54 | PHE1 |
| PHE9 | G13 | ASN26 | ARG38 | ASP34 | TYR35 | TYP207 | TYR35 |
| VAL13 | V16 | HIS78 | SER63 | ARG38 | TRP216 | | TYP216 |
| LEU73 | F228 | | CYS143 | HIS163 | | | |
| TYR103 | N236 | | | | | | |

## Conclusion

Due to non approved drugs at present Currently, COVID-19 has emerged in the human population, in China, and is a potential threat to global health, worldwide. Currently, the main target for COVID-19 treatment primarily act on the main protease (Mpro). The aim of this study was to examine Jensenone from eucalyptus essential oil that may be used to inhibit the COVID-19 infection pathway. Jensenone has

65

high binding affinity and lowest binding energies. Therefore, we suggested that may Jensenone may represent potential treatment options, and found in medicinal plants that may act as potential inhibitors of COVID-19 Mpro. However, further studies should be conducted for the validation of these compounds using in vitro and in vivo models to pave a way for these compounds in drug discovery.

## Acknowledgment

ADS want to thank management for this support.

## Conflict of interest

Authors declares no conflict of interest

## Compliance with Ethical Standards

The authors declare that they have no conflict of interest. This article does not contain any studies involving animals or human participants performed by any of the authors

## Author contributions

ADS: designed the study and prepared manuscript

IJK: designed the study and prepared manuscript

66